\documentstyle[preprint,aps,epsf,floats]{revtex}

\tighten
\preprint{\vbox{
\hbox{DOE/ER/40762-245}
\hbox{UMD-PP-02-020}
}} \bigskip \bigskip

\begin{document}
\title{Leading Chiral Contributions to the Spin Structure of the Proton}
\author{Jiunn-Wei Chen and Xiangdong Ji}
\address{Department of Physics, University of Maryland, \\
College Park, MD 20742, USA\\
{\tt jwchen@physics.umd.edu, xji@physics.umd.edu}}
\maketitle

\begin{abstract}
The leading chiral contributions to the quark and gluon components of the
proton spin are calculated using heavy-baryon chiral perturbation theory.
Similar calculations are done for the moments of the generalized parton
distributions relevant to the quark and gluon angular momentum densities.
These results provide useful insight about the role of pions in the spin
structure of the nucleon, and can serve as a guidance for extrapolating
lattice QCD calculations at large quark masses to the chiral limit.
\end{abstract}

\vfill\eject

The spin structure of the proton has been an active subject of
investigation in
nuclear and high energy physics since the EMC (European Muon Collaboration)
measurement of the $g_{1}^{p}(x,Q^{2})$ structure function in polarized muon
deep-inelastic scattering \cite{emc}. EMC found that a surprisingly small
fraction of the proton spin resides in quark spin---in flat contradiction
with the well-known SU(6) quark model prediction \cite{close}. In the last
decade, many follow-up experiments have been performed at the CERN, SLAC,
and DESY laboratories, and the next-to-leading order QCD analyses of these
data confirm the original discovery of EMC \cite{review}. The outstanding
question is then what are the other contributions to the proton spin?
According to quantum chromodynamics (QCD), the angular momentum of a
strong-interacting system can be written as a sum of quark spin (${\bf S}_{q}
$), quark orbital (${\bf L}_{q}$), and gluon angular momentum (${\bf J}_{g}$%
) contributions \cite{ji1}, 
\begin{eqnarray}
{\bf J}_{{\rm QCD}} &=&\int d^{3}x\frac{1}{2}\psi ^{\dagger }{\bf \Sigma }%
\psi +\int d^{3}x\psi ^{\dagger }{\bf x}\times \left( -i{\bf D}\right) \psi 
\nonumber \\
&&+\int d^{3}x{\bf x}\times \left( {\bf E}\times {\bf B}\right) \ ,
\end{eqnarray}
where ${\bf D}=\partial +ig{\bf A}$ is the covariant derivative, ${\bf A}$
the gluon potentials, $\psi $ the quark fields, and ${\bf E}$ and ${\bf B}$
the chromoelectric and magnetic fields. Summations over flavors and colors
are implicit. Using the above operator, one can write down a
gauge-invariant, frame-independent decomposition of the proton spin \cite
{ji1}, 
\begin{equation}
{\frac{1}{2}}={\frac{1}{2}}\Delta \Sigma (\mu ^{2})+L_{q}(\mu
^{2})+J_{g}(\mu ^{2}) \ ,
\end{equation}
in a class of frames in which the proton has a definite helicity \cite{ji2}.
Individual terms in the above equation are defined as the expectation values
of the corresponding operators in the proton helicity eigenstate. Because
these operators are not separately conserved, their matrix elements depend
on renormalization schemes and the related scale $\mu ^{2}$.

Recent analyses of the polarized deep-inelastic scattering data yield \cite
{review}, 
\begin{equation}
\Delta \Sigma(\mu^2\sim 5 {\rm GeV}^2) = 0.16\pm 0.08 \ .
\end{equation}
Meanwhile, it has been realized that the total quark and gluon contributions
to the proton spin, $J_q=\Delta\Sigma/2+L_q$ and $J_g$, can be obtained from
new sum rules involving generalized (off-forward) parton distributions \cite
{ji1}. These new distributions are found to be accessible experimentally in
deeply virtual Compton scattering \cite{ji1} or exclusive meson production 
\cite{radyushkin}. On the other hand, the matrix elements $\Delta\Sigma$, $%
L_q$ and $J_g$ can be calculated theoretically in lattice QCD and QCD sum
rule methods \cite{lattice}. In the former case, the up and down quark
masses used in the present simulations are much larger than the physical
ones, and the corresponding pion mass is larger than 500 MeV. Therefore, it
is crucial to make reliable extrapolations from the lattice data down
to the physical pion mass.

In this paper, we study the pionic contribution to the spin structure of the
proton. This contribution can be calculated because pions are Goldstone
bosons from the spontaneous breaking of the chiral symmetry present in the
QCD lagrangian. The explicit breaking of the symmetry through small up and
down quark masses induces small physical pion masses. Effective field theory
technology allows study of the symmetry breaking effects in a systematic
expansion of $m_\pi/4\pi f_\pi$. Here we focus on the leading nonanalytical
contributions of the type $m_\pi^2\ln m_\pi^2$ to the quark and gluon
components of the proton spin, $J_{q,g}$. The total nonanalytic contribution
to the proton spin, however, must vanish because the total proton spin
is independent of the pion mass.

What we actually calculate below is the chiral contribution to the proton
form factors of the quark and gluon parts of the QCD energy momentum tensor $%
T_{q,g}^{\mu \nu }$. A simple counting shows that they have three form
factors \cite{ji1,ji3}, 
\begin{eqnarray}
&&\left\langle p^{^{\prime }}\left| T_{q,g}^{\left( \mu \nu \right) }(\mu
^{2})\right| p\right\rangle =\overline{u}(p^{\prime })\left[
A_{q,g}(q^{2},\mu ^{2})\gamma ^{(\mu }\overline{p}^{\nu )}\right.   \nonumber
\\
&&\left. +B_{q,g}(q^{2},\mu ^{2})\overline{p}^{(\mu }i\sigma ^{\nu )\alpha
}q_{\alpha }/2M+C_{q,g}(q^{2},\mu ^{2})q^{(\mu }q^{\nu )}/M\right] u(p)\ ,
\end{eqnarray}
where $p=(p+p^{\prime })/2$, $q=p^{\prime }-p$, and $\left( \cdots \right) $
means that the indices enclosed are made symmetric and 
traceless. $A_{q(g)}(0,\mu
^{2})$ is the momentum fraction of the proton carried by quarks (gluons),
and therefore $A_{q}(0,\mu ^{2})+A_{g}(0,\mu ^{2})=1$. The quark (gluon)
angular momentum contribution to the proton spin is \cite{ji1} 
\begin{equation}
J_{q,g}(\mu ^{2})=\frac{1}{2}\left[ A_{q,g}(0,\mu ^{2})+B_{q,g}(0,\mu ^{2})%
\right] \ .
\end{equation}
Thus the sum rule $J_{q}(\mu ^{2})+J_{g}(\mu ^{2})=1/2$ implies $B_{q}(0,\mu
^{2})+B_{g}(0,\mu ^{2})=0$. An explicit one-loop verification of this
general result in QED can be found in Ref. \cite{brodsky}.

Since we are interested in the form factors $A(q^{2},\mu ^{2})$ and $%
B(q^{2},\mu ^{2})$ at zero momentum transfer, chiral perturbation theory is
a legitimate tool to use. As a low-energy effective field theory of QCD,
chiral perturbation theory treats chiral symmetry and the symmetry breaking
patterns of QCD in the most general fashion. In the single nucleon systems,
the baryon masses are considered heavy, and the heavy-baryon chiral
perturbation theory (HB$\chi $PT) has been formulated to 
restore the systematic power counting \cite{HBChPT}. 
In this formalism, we rewrite the form factor
in the Breit frame as 
\begin{eqnarray}
\left\langle p^{^{\prime }}\left| T_{q,g}^{\left( \mu \nu \right) }(\mu
^{2})\right| p\right\rangle  &=&\sqrt{1-\frac{q^{2}}{4M^{2}}}\overline{N}%
(v)\left\{ \left( A_{q,g}(q^{2},\mu ^{2})+B_{q,g}(q^{2},\mu ^{2})\frac{q^{2}%
}{4M^{2}}\right) Mv^{(\mu }v^{\nu )}\right.   \nonumber \\
&&+\left( A_{q,g}(q^{2},\mu ^{2})+B_{q,g}(q^{2},\mu ^{2})\right) v^{(\mu }
\left[ S^{\nu )},S\cdot q\right]   \nonumber \\
&&+\left. C_{q,g}(q^{2},\mu ^{2})q^{(\mu }q^{\nu )}/M\right\} N(v)\ ,
\end{eqnarray}
where the proton velocity $v=(p+p^{\prime })/2M$, polarization vector $%
S_{\mu }=i\gamma _{5}\sigma _{\mu \nu }v^{\nu }$ reduces to $(0,\vec{\sigma}%
/2)$ in the $v^{\mu }=(1,0,0,0)$ frame ($\overline{N}N=2M$). In the
following, we will focus
on the spin-dependent term (the second in the bracket), and compute the
chiral corrections to its coefficient, $A_{q,g}(0)+B_{q,g}(0)=2J_{q,g}$.

According to the formalism developed in Refs. \cite{cj}, the quark and gluon
energy-momentum tensor operators are ``matched'' onto a sum of hadronic
operators with the identical quantum numbers, including pure pionic
operators, single baryon operators as well as multiple baryon operators, 
\begin{equation}
T_{q,g}^{\left( \mu \nu \right) }(\mu ^{2})=T_{(q,g)\pi }^{\left( \mu \nu
\right) }(\mu ^{2},\Lambda _{\chi }^{2})+T_{(q,g)N}^{\left( \mu \nu \right)
}(\mu ^{2},\Lambda _{\chi }^{2})+T_{(q,g)\Delta }^{\left( \mu \nu \right)
}(\mu ^{2},\Lambda _{\chi }^{2})+\cdots \ ,
\end{equation}
where $\Lambda _{\chi }$ is a scale at which HB$\chi $PT is applied. Each of
these operators contains an infinite number of terms organized according
to the
HB$\chi $PT power counting. For instance, the leading terms in the pure
pionic operator are 
\begin{equation}
T_{(q,g)\pi }^{\left( \mu \nu \right) }(\mu ^{2},\Lambda _{\chi
}^{2})=a_{(q,g)\pi }(\mu ^{2})\frac{f_{\pi }^{2}}{2}\text{Tr}[\partial
^{(\mu }\Sigma ^{\dagger }\partial ^{\nu )}\Sigma ]+b_{(q,g)\pi }(\mu ^{2})%
\frac{f_{\pi }^{2}}{2}\partial ^{(\mu }\text{Tr}[\Sigma ^{\dagger }\partial
^{\nu )}\Sigma +\Sigma \partial ^{\nu )}\Sigma ^{\dagger }]+\cdots 
\end{equation}
where $\Sigma =u^{2}=\exp [i\pi ^{a}\tau ^{a}/f_{\pi }]$, $\pi ^{a}$ is the
pion field with isospin index $a$, $\tau ^{a}$ the isospin Pauli matrices, $%
f_{\pi }$ the pion decay constant $=93$ MeV. The physical meaning of $%
a_{(q,g)\pi }$ is the momentum fraction of the pion carried by quarks
(gluons) in the chiral limit. Hence, 
\begin{equation}
a_{q\pi }(\mu ^{2})+a_{g\pi }(\mu ^{2})=1\text{\ .}  \label{sr1}
\end{equation}
Experimentally, it is found that 
\begin{equation}
a_{q\pi }(\mu ^{2}\sim 1~{\rm GeV}^{2})\equiv \left\langle
x_{q}\right\rangle _{\pi }\simeq 0.5\   \label{aq}
\end{equation}
for a physical pion \cite{pion}. The leading pionic
operator contributes to the leading chiral behavior of 
the spin fractions $J_{q,g}$ shown in Fig.~1(c).

\begin{figure}[t]
\begin{center}
\epsfxsize=9.25cm
\centerline{\epsffile{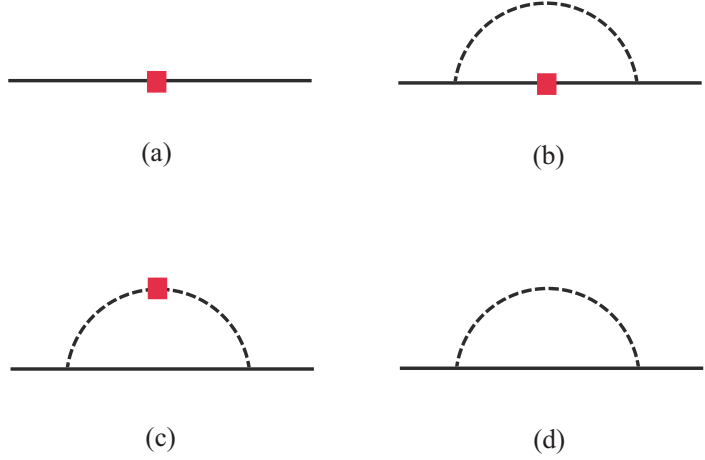}}
\end{center}
\caption{Feynman diagrams contributing to the leading chiral behavior of the
fractions of the nucleon spin carried in quarks and gluons. The dashed lines
represent pions. The diagram (d) denotes the wave function
renormalization contribution.}
\end{figure}
The leading terms in the single nucleon operator can be written as 
\begin{eqnarray}
T_{(q,g)N}^{\left( \mu \nu \right) }(\mu^2) &=&a_{(q,g)N}(\mu^2) \overline{N}%
v^{(\mu }v^{\nu )}N-b_{(q,g)N}(\mu^2)i\partial^{\alpha}\left \{ \overline{N}%
v^{(\mu }\left[ S^{\nu )},S_{\alpha }\right] N\right\}  \nonumber \\
&&+c_{(q,g)N}(\mu^2)\overline{N}{\cal A}^{(\mu }S^{\nu )}N+\cdots
\end{eqnarray}
where ${\cal A}^{\mu }=i\left( u^{\dagger }\partial ^{\mu }u-u\partial ^{\mu
}u^{\dagger }\right) =-F_{\pi }\tau ^{a}\partial ^{\mu }\pi ^{a}+\cdots $.
The second term contributes to $J_{q,g}$ through tree and one loop diagrams
(Fig. 1, diagrams (a), (b), and (d)). In fact, $b_{(q,g)N}/2$ is just $%
J_{q,g}$ in the limit where quark masses vanish 
\begin{equation}
J_{q,g}^{0}(\mu^2)=b_{(q,g)N}(\mu^2)/2\ ,  \label{dq}
\end{equation}
where the superscript $0$ indicates the chiral limit. The total nucleon
momentum and spin in the chiral limit impose the constraints 
\begin{eqnarray}
a_{qN}(\mu^2)+a_{gN}(\mu^2) &=&1\ ,  \nonumber \\
b_{qN}(\mu^2)+b_{gN}(\mu^2) &=&1\ .  \label{sr2}
\end{eqnarray}

Combining the result from different diagrams in Fig. 1, we find the
following leading chiral logarithm contributions to the spin structure of
the nucleon, 
\begin{equation}
J_{q,g}(\mu ^{2})=\frac{1}{2}\left\{ b_{(q,g)N}(\mu ^{2})+3[a_{(q,g)\pi
}(\mu ^{2})-b_{(q,g)N}(\mu ^{2})]{\frac{g_{A}^{2}m_{\pi }^{2}}{(4\pi f_{\pi
})^{2}}}\ln \left( {\frac{m_{\pi }^{2}}{\Lambda _{\chi }^{2}}}\right)
\right\} +\cdots \ ,
\end{equation}
This is the main result of this paper. The chiral
logarithms in $J_{q}(\mu ^{2})+J_{g}(\mu ^{2})$ cancel because of the
constraints in Eqs.~(\ref{sr1},~\ref{sr2}).

The delta resonance plays a special role in the nucleon structure physics.
In a world where the number of quark colors ($N_{c}$) go to infinity,
the
delta resonance is degenerate with the nucleon, i.e. $\Delta =M_{\Delta
}-M_{N}={\cal O}(1/N_{c})$. If, at the same time, the standard spontaneous
chiral symmetry breaking occurs, the chiral corrections to the nucleon
properties are strongly affected by the degenerate delta contribution. In
the real world where $\Delta \sim 300$ MeV, the significance of the delta
contribution depends strongly on the spin-isospin channel to which an
observable belongs \cite{cohen}. It can be calculated in an effective field
theory approach in which $\Delta $ is counted as the same order as $m_{\pi
}$ 
\cite{HBChPT,hammert}.

\begin{figure}[t]
\begin{center}
\epsfxsize=15.25cm
\centerline{\epsffile{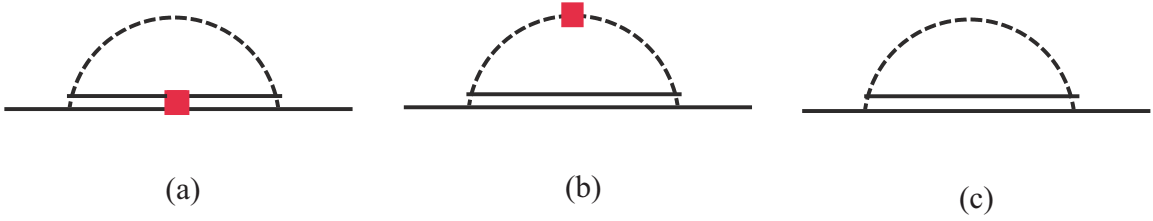}}
\end{center}
\caption{Same as Fig. 1. Contributions from the delta intermediate
states (double lines). The diagram (c) denotes the wave function
renormalization contribution.}
\label{fig:1}
\end{figure}

In Fig.~2, we have shown three different possible ways that the delta
resonance can contribute as an intermediate state to the spin structure of
the nucleon: First, the delta is present when the energy-momentum tensor of
the pion contributes (Fig. 2b); second, the delta contributes to the wave
function renormalization of the nucleon (Fig. 2c); and finally the
energy-momentum tensor of the delta contributes directly (Fig. 2a) 
\begin{equation}
T_{\Delta _{q,g}}^{\left( \mu \nu \right) }(\mu ^{2})=-3b_{(q,g)\Delta }(\mu
^{2})i\partial ^{\alpha }\left\{ \overline{\Delta }^{\beta }v^{(\mu }\left[
S^{\nu )},S_{\alpha }\right] \Delta _{\beta }\right\} +...
\end{equation}
Here the leading term contributes to the delta spin in the chiral limit.
Indeed, the total spin of the delta leads to the following constraint, 
\begin{equation}
b_{q\Delta }(\mu ^{2})+b_{g\Delta }(\mu ^{2})=1\ .  \label{sr3}
\end{equation}

The calculation of the delta contribution to $J_{q,g}(\mu ^{2})$ is
straightforward. The combined leading non-analytic chiral correction is 
\begin{eqnarray}
J_{q,g}&&(\mu ^{2}) =\frac{1}{2}\left\{ b_{(q,g)N}(\mu ^{2})+3[a_{(q,g)\pi
}(\mu ^{2})-b_{(q,g)N}(\mu ^{2})]{\frac{g_{A}^{2}m_{\pi }^{2}}{(4\pi f_{\pi
})^{2}}}\ln \left( {\frac{m_{\pi }^{2}}{\Lambda _{\chi }^{2}}}\right)
\right.   \nonumber \\
&&\left. -\left( \frac{9}{2}b_{(q,g)N}(\mu ^{2})+3a_{(q,g)\pi }(\mu ^{2})-%
\frac{15}{2}b_{(q,g)\Delta }(\mu ^{2})\right) {\frac{(2\sqrt{2}g_{\pi
N\Delta })^{2}}{(3\cdot 4\pi f_{\pi })^{2}}}K(m_{\pi },\Delta)\right\} +\cdots \ ,  \label{result}
\end{eqnarray}
where 
\begin{equation}
K(m_{\pi },\Delta)=\left( m_{\pi }^{2}-2\Delta ^{2}\right)
\log \left( {\frac{m_{\pi }^{2}}{\Lambda _{\chi }^{2}}}\right) +2\Delta 
\sqrt{\Delta ^{2}-m_{\pi }^{2}}\log \left( {\frac{\Delta -\sqrt{\Delta
^{2}-m_{\pi }^{2}+i\epsilon }}{\Delta +\sqrt{\Delta ^{2}-m_{\pi
}^{2}+i\epsilon }}}\right) \ .
\end{equation}
Again, all the delta contributions cancel when summed over quark and gluon
operators because of the constraints on the low-energy constants.

Another nontrivial check of the above result is to let the number
of colors $N_{c}$ go to infinity. In this limit, the individual nucleon and
delta contributions scale as $N_{c}$. However, because $2\sqrt{2}g_{\pi
N\Delta }/3=g_{A}$ and $b_{(q,g)N}=b_{(q,g)\Delta }$ at large $N_c$
\cite{CJ1}, the order $N_{c}$ chiral contributions cancel. In the 
real world, we can use the large-$N_{c}$ approximation
of the delta observables to get an estimate of the total
chiral contributions, 
\begin{equation}
J_{q,g}(\mu ^{2})=J_{q}^{0}+{\frac{3g_{A}^{2}}{(4\pi f_{\pi })^{2}}}\left(
J_{q}^{0}-\frac{\left\langle x\right\rangle _{q/\pi }}{2}\ \right) \left[
K(m_{\pi },\Delta)-m_{\pi }^{2}\ln \frac{m_{\pi }^{2}}{%
\Lambda _{\chi }^{2}}\right] ,
\end{equation}
which depends on the unknown $J_{q}^{0}$. It is interesting to 
note that the non-analytic corrections will be largely 
cancelled if $J_{q}^{0}\approx
\left\langle x\right\rangle _{q/\pi }/2$, a possibility favored
by empirical considerations. In this case, a linear $m_\pi^2$ 
extrapolation of lattice calculations 
to the chiral limit might be justified.

In Refs. \cite{hoodbhoy}, the quark and gluon angular momentum distributions
are found to be $J_{q}(x)=x(\Sigma (x)+E_{q}(x,0,0))/2$ and $%
J_{g}(x)=x(g(x)+E_{g}(x,0,0))/2$, where $\Sigma (x)$ and $g(x)$ are
unpolarized (singlet) quark and gluon distributions and $E_{q,g}(x,0,0)$ are
generalized parton distributions \cite{gpd}. The moments of these distributions are
related to the form factors of the twist-two operators \cite{ji3}. 
In leading order in $q$, 
\begin{eqnarray}
\langle p^{\prime }|{\cal O}^{\mu _{1}\mu _{2}...\mu _{n}}|p\rangle 
&=&A_{n,0}(0,\mu ^{2}){\overline{N}}(v)v^{\mu _{1}}v^{\mu _{2}}...v^{\mu
_{n}}N(v)M^{n-1}  \nonumber \\
&&+[A_{n,0}(0,\mu ^{2})+B_{n,0}(0,\mu ^{2})]{\overline{N}}(v)[S^{\mu
_{1}},S\cdot q]v^{\mu
_{2}}...v^{\mu _{n}}N(v)M^{n-2} +... \ .
\end{eqnarray}
Then the relation is 
\begin{equation}
\int_{-1}^{1}x^{n-2}J_{q,g}(x,\mu ^{2})dx={\frac{1}{2}}[A_{n,0}(0,\mu
^{2})+B_{n,0}(0,\mu ^{2})] \ ,
\end{equation}
for $n=2,4,...$. By going through the same calculation as above, one can
show that, 
\begin{eqnarray}
\int_{-1}^{1}x^{n-2}&&J_{q,g}(x,\mu ^{2})dx =b_{n(q,g)N}(\mu ^{2})\left( 1-3{%
\frac{g_{A}^{2}m_{\pi }^{2}}{(4\pi f_{\pi })^{2}}}\ln \left( {\frac{m_{\pi
}^{2}}{\Lambda _{\chi }^{2}}}\right) \right)   \nonumber \\
&&-\left( \frac{9}{2}b_{n(q,g)N}(\mu ^{2})-\frac{15}{2}b_{(q,g)\Delta }(\mu
^{2})\right) {\frac{(2\sqrt{2}g_{\pi N\Delta })^{2}}{(3\cdot 4\pi f_{\pi
})^{2}}}K(m_{\pi },\Delta)+\cdots \ ,
\end{eqnarray}
for $n=4,6,...$, where $b_{n(q,g)N}$ and $b_{n(q,g)\Delta }$ are the
coefficients of the nucleon and delta spin-dependent twist two operators in
the effective theory. Apart from the absence of the pion contribution, the
other
terms are similar to the $n=2$ moment.

To summarize, we have calculated the leading chiral contribution to the spin
structure of the nucleon. These results provide interesting
insight about the role of the pion in the composition of the 
nucleon spin. It also provides useful guidance for extrapolating 
lattice QCD calculations at large quark masses to the 
chiral limit \cite{lattice,thomas}.

This work is supported in part by the U.S. Dept. of Energy under grant No.
DE-FG02-93ER-40762-245.


\begin{references}
\bibitem{emc}  J. Ashman {\it et al.,} Phys. Lett. B 202 (1988) 603; Nucl.
Phys. B 328 (1989) 1.

\bibitem{close}  F. E. Close, {\em An Introduction to Quarks and Partons},
Academic Press, london (1979). 

\bibitem{review}  For a review, see 
B. W. Filippone and X. Ji, hep-ph/0101224, Adv. in Nucl. Phys. 
Vol. 26, 1 (2001). 

\bibitem{ji1}  X. Ji, Phys. Rev. Lett. 78, 610 (1997); Phys. Rev. D 55, 7114
(1997).

\bibitem{ji2}  X. Ji, Phys. Rev. D 58, 056003 (1998).

\bibitem{radyushkin}  A. V. Radyushkin, Phys. Lett. B 385, 333 (1996); J. C.
Collins, L. Frankfurt, and M. Strikman, Phys. Rev. D 56, 2982 (1997).

\bibitem{lattice}  N. Mathur, S. J. Dong, K. F. Liu, L. Mankiewicz, and N.
C. Mukhopadhyay, Phys. Rev. D 62, 114504 (2000); V. Gadyiak, X. Ji, and C.
Chung, to be published; X. Ji and I. Balitsky, Phys. Rev. Lett. 79, 1225
(1997).

\bibitem{ji3}  X. Ji and R. Lebed, Phys. Rev. D 63, 076005 (2001).

\bibitem{brodsky}  S. J. Brodsky, D. Hwang, B.-Q. Ma, I. Schmidt, Nucl.
Phys. B 593, 311 (2001).

\bibitem{HBChPT}  E. Jenkins and A. V. Manohar, Phys. Lett. B 255, 558
(1991).

\bibitem{cj}  J. W. Chen and X. Ji, hep-ph/0105197; D. Arndt and M. J.
Savage, nucl-th/0105045; See also, J. W. Chen and X. Ji, Phys. Rev. Lett.
87, 152002 (2001). 

\bibitem{pion}  M. Gluck, E. Reya and A. Vogt Z. Phys. C53, 651 (1992); M.
Gluck, E. Reya and I. Schienbein, Eur. Phys. J. C10, 313 (1999). 

\bibitem{cohen}  T. Cohen, Rev. Mod. Phys. 68, 599 (1996). 

\bibitem{hammert}  T.R. Hemmert, B.R. Holstein and J. Kambor J. Phys. G 24,
1831 (1998). 

\bibitem{CJ1}  J. W. Chen and X. Ji, hep-ph/0105296. 

\bibitem{hoodbhoy}  P. Hoodbhoy, X. Ji and W. Lu, Phys. Rev. D 59, 014013
(1999); Phys. Rev. D 59, 074010 (1999).  

\bibitem{gpd}  See for example, X. Ji, J. Phys. G {\bf 24}, 1181 (1998); A.
V. Radyushkin, hep-ph/0101225; K. Goeke, M. V. Polyakov, M. Vanderhaeghen,
hep-ph/0106012.


\bibitem{thomas}  W. Detmold, W. Melnitchouk, J. W. Negele, D. B. Renner,
and A. W. Thomas, hep-lat/0103006; A. W. Thomas, W. Melnitchouk, F. M.
Steffens, Phys. Rev. Lett. {\bf 85}, 2892 (2000).
\end{references}
\end{document}